\documentclass[iop,apjl]{emulateapj}

\usepackage{natbib}
\slugcomment{To appear in the Astrophysical Journal}

\shorttitle{Elemental abundances of \cyga\ and B}
\shortauthors{Ram\'irez et~al.}

\newcommand{\feh}{\mathrm{[Fe/H]}}
\newcommand{\teff}{T_\mathrm{eff}}
\newcommand{\logg}{\log g}

\newcommand{\fei}{Fe\,\textsc{i}}
\newcommand{\feii}{Fe\,\textsc{ii}}

\newcommand{\kms}{km\,s$^{-1}$}
\newcommand{\tc}{T_\mathrm{C}}
\newcommand{\vt}{v_t}
\newcommand{\cyga}{16\,Cyg\,A}
\newcommand{\cygb}{16\,Cyg\,B}
\newcommand{\mgib}{Mg\,\textsc{i}\,$b$}

\begin{document}

\title{Elemental abundance differences in the 16\,Cygni binary system: \\
a signature of gas giant planet formation?}

\author{I.\,Ram\'irez\altaffilmark{1},
        J.\,Mel\'endez\altaffilmark{2},
        D.\,Cornejo\altaffilmark{3},
        I.\,U.\,Roederer\altaffilmark{1}, and
        J.\,R.\,Fish\altaffilmark{1,4}
        }

\email{ivan@obs.carnegiescience.edu}
\altaffiltext{1}{The Observatories of the Carnegie Institution for Science;
                 813 Santa Barbara Street, Pasadena, CA 91101, USA}
\altaffiltext{2}{Departamento de Astronomia do IAG/USP, Universidade de S\~ao Paulo; Rua do M\~atao 1226, S\~ao Paulo, 05508-900, SP, Brasil}
\altaffiltext{3}{Departamento de Astrof\'isica, Agencia Espacial del Per\'u CONIDA; Luis Felipe Villar\'an 1069, San Isidro, Lima, Per\'u}
\altaffiltext{4}{Harvey Mudd College; 301 Platt Boulevard, Claremont, CA 91711, USA}

\begin{abstract}
The atmospheric parameters of the components of the 16\,Cygni binary system, in which the secondary has a gas giant planet detected, are measured accurately using high quality observational data. Abundances relative to solar are obtained for 25 elements with a mean error of $\sigma(\mathrm{[X/H]})=0.023$\,dex. The fact that \cyga\ has about four times more lithium than \cygb\ is normal considering the slightly different masses of the stars. The abundance patterns of \cyga\ and B, relative to iron, are typical of that observed in most of the so-called solar twin stars, with the exception of the heavy elements $(Z>30)$, which can, however, be explained by Galactic chemical evolution. Differential (A--B) abundances are measured with even higher precision ($\sigma(\Delta\mathrm{[X/H]})=0.018$\,dex, on average). We find that \cyga\ is more metal-rich than \cygb\ by $\Delta\mathrm{[M/H]}=+0.041\pm0.007$\,dex. On an element-to-element basis, no correlation between the A--B abundance differences and dust condensation temperature ($\tc$) is detected. Based on these results, we conclude that if the process of planet formation around \cygb\ is responsible for the observed abundance pattern, the formation of gas giants produces a constant downwards shift in the photospheric abundance of metals, without a $\tc$ correlation. The latter would be produced by the formation of terrestrial planets instead, as suggested by other recent works on precise elemental abundances. Nevertheless, a scenario consistent with these observations requires the convective envelopes of $\simeq1M_\odot$ stars to reach their present-day sizes about three times quicker than predicted by standard stellar evolution models.
\end{abstract}


\section{Introduction}

It is often assumed that the members of multiple stellar systems with similar components have the exact same chemical composition since they formed from a common molecular cloud and have undergone similar evolutionary paths. A number of studies have found, however, that elemental abundance differences in binary stars are not uncommon \citep[e.g.,][]{gratton01,laws01,sadakane03,desidera04,desidera06}. Systematic errors in the spectroscopic analysis could be blamed for some of these anomalies but small differences have been observed even in pairs with very similar components, i.e., in cases where differential analysis is expected to minimize the errors. The origin of these abundance anomalies remains unknown.

Stars and their planets form nearly at the same time. Therefore, it is likely that the process of planet formation leaves detectable signatures in the chemical composition of the host stars. The well-known planet-metallicity correlation in late-type main-sequence stars, for example, suggests that high metallicity increases the probability of forming planets, at least the gas giants \citep[e.g.,][]{marcy05,udry07,johnson10}. \citet[][hereafter M09]{melendez09:twins} have further suggested that small anomalies detected in the solar chemical composition compared to that of the majority of so-called solar twin stars, objects with spectra nearly indistinguishable from the solar spectrum, could be due to the formation of terrestrial planets \citep[see also][hereafter R09]{ramirez09}. They detect a small deficiency of refractory elements in the solar photosphere which would disappear if the rocky material that once formed in the solar system were diluted into the present-day solar convective envelope \citep{chambers10}. Thus, the missing mass of refractories in the Sun seems to have been locked-up in the rocky bodies of the solar system, including the terrestrial planets.

The hypothesis described above assumes that the formation of the massive gas giants in the solar system did not affect the solar chemical composition, which seems unlikely given the much larger masses of these planets in comparison with the terrestrial planets. Nevertheless, this could be the case if, for example, the abundance ratios of different metals in the gas giants combined were similar to those found in the Sun. Unfortunately, the detailed chemical composition of the gas giants, in particular their cores, is not yet known with enough precision to rule out or confirm this assumption. Careful spectroscopic studies of stars with gas giant planets detected could therefore provide us with important clues about the impact of planet formation on the chemical composition of the host stars.

The binary star 16\,Cygni is an ideal system to look for small elemental abundance anomalies and to study the impact of gas giant planet formation. 16\,Cyg has long been known as the nearest bound pair of solar twin stars \cite[e.g.,][]{friel93,porto97,hauser99}, although strictly speaking they should be referred to as solar analogs \cite[cf.][]{melendez07:twins}, in particular because they appear to be older than the Sun. Careful line-by-line differential analysis of this type of stars yields elemental abundances with the highest precision possible, as demonstrated by the works of M09 and R09. Moreover, the components of the 16\,Cyg system are bright ($V\sim6$\,mag) and therefore very high quality spectra can be obtained for them using mid-sized telescopes. Thus, extremely high precision differential spectroscopic analysis can be applied to this system.

The secondary star in the 16\,Cyg binary system, \cygb, is known to host a gas giant planet with a minimum mass of 1.5 Jupiter masses (1.5\,$M_\mathrm{J}$) in a highly eccentric ($e\simeq0.6$) orbit. \cite{cochran97} were the first to detect this planet and they also obtained accurate radial velocity data that showed that \cyga\ does not host the same type of planet (gas giant in short-period orbit). Up-to-date follow-up observations of this pair of stars have not changed their planetary status (W.~Cochran, private communication). The presence of low-mass and/or long orbital period planets around either of these two stars is, of course, not to be ruled out.

A faint companion, likely a cool M-dwarf, to the primary star \cyga, at an angular distance of about 3.4\,arcsec has been reported by a number of authors \citep[e.g.,][]{hauser99,turner01,patience02}. The absolute magnitude difference between \cyga\ and C (the faint companion) is about 8.6 mag in the $R$ band. Their reported separation is large enough for the light from 16\,Cyg\,C to not interfere with that from \cyga\ in our data but even if 16\,Cyg\,C was superposed with \cyga\ at the time of our observation, it would affect the combined spectrum by less than 0.04\,\% at $\lambda=6000$\,\AA\ and even less at shorter wavelengths.

The atmospheric parameters of the 16\,Cyg pair have been measured by a number of authors using a variety of techniques. Few of these previous studies, however, present results based on extremely high precision analysis, i.e., significantly better than ``standard'' methods. One of these studies is that by \cite{laws01}, who find that \cyga\ is more iron-rich than \cygb\ by about $6\pm2$\,\%. They also comment as a note added in proof that a significant positive correlation with dust condensation temperature is found for the abundance difference of 13 elements analyzed. They claim that their results support a self-pollution model in which \cyga\ accreted planetary material that once formed there. They also argue that this scenario could explain the observed difference in lithium abundance between \cyga\ and B \cite[e.g.,][]{king97}. Later work by \cite{takeda05}, however, suggests that \cyga\ and B have exactly the same chemical composition and attributed the discrepancy with the results by \citeauthor{laws01} to differences in the adopted stellar parameters.

Here we derive atmospheric parameters and elemental abundances as precisely and accurately as possible using high quality data of \cyga\ and B. High precision elemental abundances are presented for 25 elements in the 16\,Cyg binary system. We discuss scenarios that could explain the observed abundance patterns.

\section{Data} \label{s:data}

The photometric data used in this paper, i.e., magnitudes and colors, correspond to those listed in the General Catalogue of Photometric Data \cite[GCPD,][]{mermilliod97}. We adopted the mean values (and errors) given in the GCPD, which have been computed using data from several different sources. For the $(V,B-V)$ pair, for example, more than 40 observations from about 10 sources are available for both \cyga\ and B. 

To calculate the distance to the 16\,Cyg system, we adopted the measured trigonometric parallaxes from the latest reduction of Hipparcos data \citep{vanleeuwen07}. Averaging the published values for the two stars we derive a parallax of $47.29\pm0.21$\,mas, which corresponds to a distance of $21.15\pm0.09$\,pc.

High resolution ($R=\lambda/\Delta\lambda\simeq60,000$), high signal-to-noise ratio ($S/N\simeq300-500$, per pixel) spectra were obtained by us with the R.\,G.\,Tull coud\'e spectrograph on the 2.7\,m Harlan~J.\,Smith Telescope at McDonald Observatory on April 25, 2011. These spectra cover the wavelength region from 4130 to 10\,000\,\AA, complete up to about 5935\,\AA, and with increasingly wider gaps of 2 to 120\,\AA\ towards redder wavelengths. The spectral windows from 6360 to 6440\,\AA\ and from 6480 to 6520\,\AA\ are affected by the so-called ``picket fence,'' a straylight source of emission lines within the spectrograph \citep[see Sect.~4.6 in ][]{tull95}, and therefore were not used in this work. Two exposures of 20 minutes each were acquired for both \cyga\ and B, one immediately after the other. Data were reduced in the standard manner with the IRAF \verb"echelle" package, using identical reduction parameters for the two stars (e.g., polynomial orders for the overscan and scattered light removal as well as order tracing and aperture widths, flat fields, wavelength solution, etc.). The spectrograph was set so that the H$\alpha$ line at 6562\,\AA\ fell in the center of the detector. This choice was made to properly normalize the H$\alpha$ line profile for effective temperature determination (Sect.~\ref{s:halpha}). Continuum normalization was done by fitting high order polynomials to the upper envelopes of each spectral order. Overlapping regions from continuous orders were later merged using the observed counts as weights. The order containing the H$\alpha$ line as well as that containing the \mgib\ lines were normalized differently. For these orders, we took advantage of the smooth variation of the blaze function from one spectral order to the next, to perform a 2D continuum normalization (i.e., both parallel and perpendicular to the direction of dispersion) excluding orders affected by strong lines, as described in \cite{barklem02}.

Daysky spectra were acquired in the afternoon of April 24, 2011 using the same spectrograph but letting the sky light enter the slit from a ``solar port'' which points towards the zenith. The data reduction of these spectra, however, was different from that of the stars. Daysky fills up the slit as the source is not point-like. In addition, scattered light in the Earth's atmosphere is known to affect the line depths and also equivalent widths \citep[e.g.,][]{gray00}. The latter effect is stronger for large angular separations between the Sun and the area of the sky observed and minimized when the area of the sky observed is very close to the location of the Sun.

\begin{figure}
\includegraphics[bb=70 365 390 685,width=8.7cm]{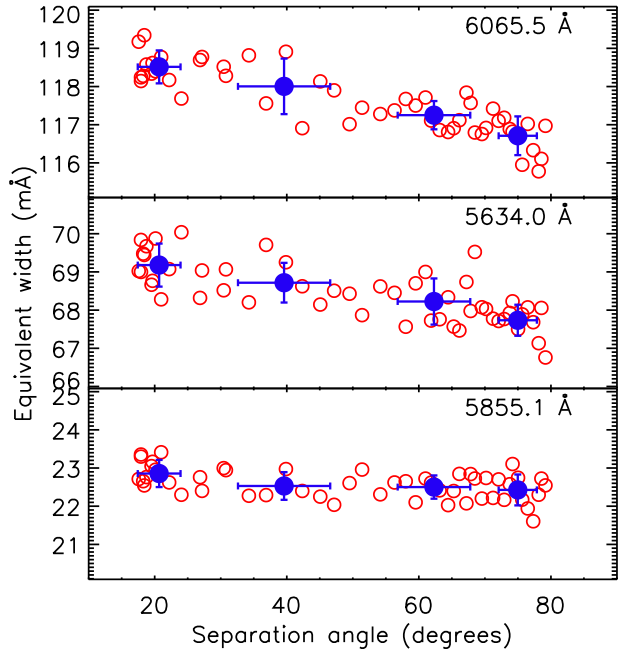}
\caption{Equivalent widths measured in our daysky spectra as a function of separation angle for three representative lines (open circles). Filled circles with error bars correspond to bin-averaged values.}
\label{f:daysky1}
\end{figure}

\begin{figure}
\includegraphics[bb=75 370 560 702,width=8.9cm]{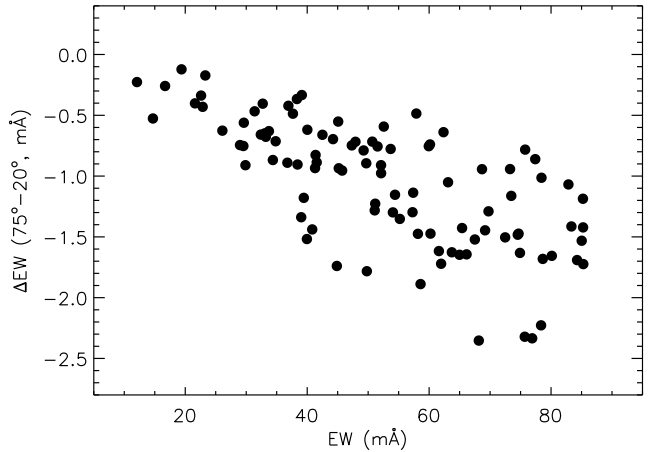}
\caption{Equivalent width difference measured in our solar spectra at 75 degrees minus 20 degrees of separation angle as a function of the equivalent width measured at 20 degrees of separation angle. Bin-averaged $EW$s (cf. Fig.~\ref{f:daysky1}) were used in this figure.}
\label{f:daysky2}
\end{figure}

We obtained daysky spectra for zenith--Sun separation angles from 20 to 80 degrees. The measured equivalent widths clearly correlate with the separation angle (Fig.~\ref{f:daysky1}) while the angular variations seem more important for stronger features (Fig.~\ref{f:daysky2}). Our daysky spectra reveal the systematic effects described in \cite{gray00} and are thus not the best reference for very high precision abundance work. We obtained a number of daysky spectra when the Sun was at its highest point in the sky, about 20 degrees from zenith. Although not ideal, the problems described above are minimized under these circumstances. Thus, our solar reference spectrum corresponds to the average of 10 daysky spectra taken with the Sun near maximum height. We stress that the results obtained using our solar spectrum as reference are expected to be less precise than a differential analysis of \cyga\ relative to \cygb, since the data acquisition and reduction process for the latter are more consistent.

We can use the measurements from Fig.~\ref{f:daysky1} to estimate the error in our measured line equivalent widths ($EW$s), since the $S/N$ of each of our solar spectra is comparable to that of our \cyga\ and B spectra. On average, the error is $\simeq0.5$\,m\AA, which corresponds to about 1\,\% for a typical line used in this work ($EW\simeq50$\,m\AA).

\begin{table*}
\caption{Adopted atomic data and our measured equivalent widths.}
\label{t:lines}
\centering
\begin{tabular}{ccccc||ccc}\hline\hline
$\lambda$ (\AA) & species & EP (eV) & $\log(gf)$ & source & EW (m\AA) & EW (m\AA) & EW (m\AA) \\
 & & & & & \cyga & \cygb & Sun \\ \hline 
 5052.1670 &    C & 7.685 & -1.304 & A09 &  40.0 &  35.7 &  34.7 \\
 5380.3369 &    C & 7.685 & -1.615 & A09 &  27.2 &  23.8 &  22.7 \\
 7771.9438 &    O & 9.146 &  0.352 & A09 &  82.6 &  74.8 &  71.9 \\
 7774.1611 &    O & 9.146 &  0.223 & A09 &  71.5 &  65.5 &  62.5 \\
 7775.3901 &    O & 9.146 &  0.002 & A09 &  57.1 &  51.7 &  48.6 \\
 6160.7471 &   Na & 2.104 & -1.246 & A09 &  61.2 &  61.5 &  55.9 \\
 4730.0400 &   Mg & 4.340 & -2.390 & R03 &  75.9 &  76.3 &  69.1 \\
 5711.0879 &   Mg & 4.345 & -1.729 & A09 & 111.6 & 112.3 & 107.8 \\
 6318.7168 &   Mg & 5.108 & -1.945 & M09 &  45.2 &  44.4 &  43.7 \\
 6319.2368 &   Mg & 5.108 & -2.324 & A09 &  31.6 &  30.1 &  25.0 \\
 6696.0181 &   Al & 3.143 & -1.481 & A09 &  48.4 &  47.9 &  40.5 \\
 6698.6670 &   Al & 3.143 & -1.782 & A09 &  25.9 &  26.7 &  21.2 \\
 5665.5542 &   Si & 4.920 & -1.940 & M09 &  48.6 &  47.7 &  41.9 \\
 5517.5400 &   Si & 5.080 & -2.496 & N09 &  17.6 &  16.6 &  13.7 \\
 5690.4248 &   Si & 4.930 & -1.773 & A09 &  60.0 &  58.0 &  52.0 \\
 5701.1050 &   Si & 4.930 & -1.953 & A09 &  46.0 &  44.8 &  38.8 \\
 5772.1450 &   Si & 5.082 & -1.653 & A09 &  60.2 &  58.1 &  53.8 \\
 5793.0732 &   Si & 4.930 & -1.963 & A09 &  52.2 &  50.1 &  45.6 \\
 5948.5400 &   Si & 5.080 & -1.208 & N09 &  94.5 &  92.6 &  90.0 \\
 6125.0298 &   Si & 5.610 & -1.510 & R03 &  39.3 &  37.5 &  32.8 \\
 6142.4902 &   Si & 5.620 & -1.540 & R03 &  38.3 &  37.5 &  32.3 \\
 6145.0200 &   Si & 5.610 & -1.480 & R03 &  47.8 &  45.1 &  39.6 \\
 6237.3301 &   Si & 5.610 & -1.116 & N09 &  73.6 &  71.4 &  65.1 \\
 6244.4800 &   Si & 5.610 & -1.360 & R03 &  55.7 &  54.7 &  48.1 \\
 6721.8398 &   Si & 5.863 & -1.060 & R03 &  55.7 &  54.0 &  47.4 \\
 7405.7700 &   Si & 5.614 & -0.720 & B03 & 101.4 &  99.2 &  94.8 \\
 7800.0000 &   Si & 6.180 & -0.710 & R03 &  64.7 &  60.1 &  58.6 \\
 4695.4429 &    S & 6.525 & -1.829 & A09 &   9.5 &   8.2 &   7.1 \\
 6743.5400 &    S & 7.866 & -0.600 & M09 &  10.1 &   8.4 &   8.3 \\

$\ldots$ & $\ldots$ & $\ldots$ & $\ldots$ & $\ldots$ & $\ldots$ & $\ldots$ & $\ldots$ \\ \hline
\end{tabular}
\end{table*}

We compiled a long list of spectral lines for the elemental abundance determination. Due to the differential nature of our work, and the similarity of the two brightest stars in the 16\,Cyg system, as well as their similarity to the Sun, accurate atomic data are not strictly necessary. We started with the linelist used by \citet[hereafter A09]{asplund09:review}, which was carefully constructed for their solar abundance analysis. Not only are these lines mostly free from severe blends but also their atomic data are accurately known. We complemented the iron linelist with the compilation by \citet[hereafter R07]{ramirez07}, which consists of lines that appear mostly unblended in solar-type stars and have reasonably accurate atomic data measured in the laboratory. For the rest of elements studied in this paper, we used additional lines listed in other large studies of solar-type stars: \citet[hereafter R03]{reddy03}, \citet[hereafter B05]{bensby05}, \citet[hereafter N09]{neves09}, and M09. Our adopted linelist is given in Table~\ref{t:lines}, along with the source from which atomic data were adopted. When available, we used the van der Waals damping constants by \cite{barklem00} and \cite{barklem05}, otherwise we adopted the values obtained from Uns\"old's approximation but with the $C_6$ constants \cite[e.g.,][page~217]{gray92:book} multiplied by a factor of 3.1, a number that we estimated empirically from a study of the lines that are included in Barklem's work.

Spectral line equivalent widths ($EW$s) were measured interactively using the \verb"splot" task in IRAF. The locations of the local continuum, which are crucial for precise measurements of $EW$, were selected as carefully as possible. For each line, we searched for regions clearly free from spectral lines in the three spectra: solar, \cyga\ and B. The exact same wavelengths for the continuum were used for all three spectra. If a small blend appeared to affect only one of the line wings, the other wing was used to measure $EW$ by fitting only that side of the spectral line to a Gaussian profile. This procedure tends to result in more precise $EW$ values than deblending the line because of the uncertainty in determining the core wavelength of the blend. In these cases, we also ensured that the local continua for all three spectra were chosen as similarly as possible. We excluded features affected by telluric absorption.

Equivalent width analysis allowed us to measure abundances of 22 elements: C, O, Na, Mg, Al, Si, S, K, Ca, Sc, Ti, V, Cr, Mn, Fe, Co, Ni, Cu, Zn, Y, Zr, and Ba but we also measured abundances of the heavy elements Nd and Eu. Most of the Nd\,\textsc{ii} and Eu\,\textsc{ii} features are very weak and heavily blended in solar-type spectra. Thus, spectrum synthesis was necessary to measure their abundances. To ensure high precision, we restricted our analysis to the best lines available, i.e., those with a very good continuum normalization and clearly identified blends with proper modeling and atomic parameters. We analyzed 2 Eu\,\textsc{ii} lines (4205 and 6645.10\,\AA) and 3 Nd\,\textsc{ii} lines (5293.16, 5319.81, and 5811.57\,\AA). For Eu we adopted the following isotopic fractions: $^{151}$Eu$=0.478$ and $^{153}$Eu$=0.522$ \citep[e.g.,][]{lodders03}. The relevant laboratory data used (transition probabilities, isotopic shifts, etc.) are given by \cite{denhartog03}, \cite{lawler01}, \cite{ivans06}, and \cite{roederer08}. Finally, lithium abundances were derived using line synthesis of the 6708\,\AA\ Li doublet, as described in \cite{baumann10}.

\section{Atmospheric parameters}

The results presented in this section were obtained using the grid of MARCS model atmospheres \citep{gustafsson08} available online.\footnote{http://marcs.astro.uu.se} Tests made with different flavors of Kurucz' model atmospheres show that the choice of atmospheric models has a very minor impact on our derived relative parameters and abundances. We comment briefly on these effects in later sections of this paper. Curve-of-growth analysis was used to measure elemental abundances from our equivalent width measurements using the 2010 version of the spectrum synthesis code MOOG \citep{sneden73}.\footnote{http://www.as.utexas.edu/$\sim$chris/moog.html} Line profile fitting was used to better constrain $\teff$ from the H$\alpha$ line and $\logg$ from the \mgib\ lines. For the former, we used a fine grid of H$\alpha$ theoretical line profiles (steps of 10\,K in $\teff$, and 0.05\,dex in $\logg$ and $\feh$), computed as in \cite{barklem02}.\footnote{These H$\alpha$ profiles were also computed using MARCS atmospheric models.} Finally, the synthesis of the \mgib\ lines was done using SYNSPEC \citep{hubeny88,hubeny95}.\footnote{http://nova.astro.umd.edu/Synspec43/synspec.html}

Stellar parameters often depend on each other so the process of their determination is intrinsically iterative. Here we present results which are self-consistent, in the sense that when we ``adopt'' a value below in the calculation of a given parameter, we have made sure that no further changes to any of the other parameters involved will occur after another iteration.

\subsection{Excitation/ionization equilibrium}

The iron lines alone can be used to determine the atmospheric parameters using the conditions of excitation and ionization equilibrium. We start by assuming that the atmospheric parameters $\teff$ (effective temperature), $\logg$ (surface gravity), $\feh$ (iron abundance),\footnote{We use the standard notation: $A_\mathrm{X}=\log(n_\mathrm{X}/n_\mathrm{H})+12$, where $n_\mathrm{X}$ is the number density of element X, and $\mathrm{[X/H]}=A_\mathrm{X}-A_\mathrm{X}^\odot$.} and $v_t$ (microturbulent velocity) of \cyga\ and B are solar (i.e., $\teff,\logg,\feh,\vt=5777$\,K$,4.437,0.00,1.00$\,\kms).\footnote{The solar microturbulent velocity $v_t$ is somewhat uncertain and dependent on the iron line list used and model atmosphere adopted. Its exact value is not crucial for our strictly differential study, as shown later.} Then, $\teff$ and $v_t$ were iteratively modified so that trends of iron abundance (from \fei\ lines only) with excitation potential (EP) and reduced equivalent width ($\log EW/\lambda$), respectively, approached zero. Simultaneously, $\logg$ was modified so that the mean difference in iron abundance from \fei\ and \feii\ lines also approached zero. Errors in the derived parameters were computed from the uncertainty in the measured slope of abundance versus EP (for $\teff$) and from the standard deviation of the mean difference of \fei\ and \feii\ abundances (for $\logg$).

\begin{figure}
\includegraphics[bb=60 360 560 712,width=8.7cm]{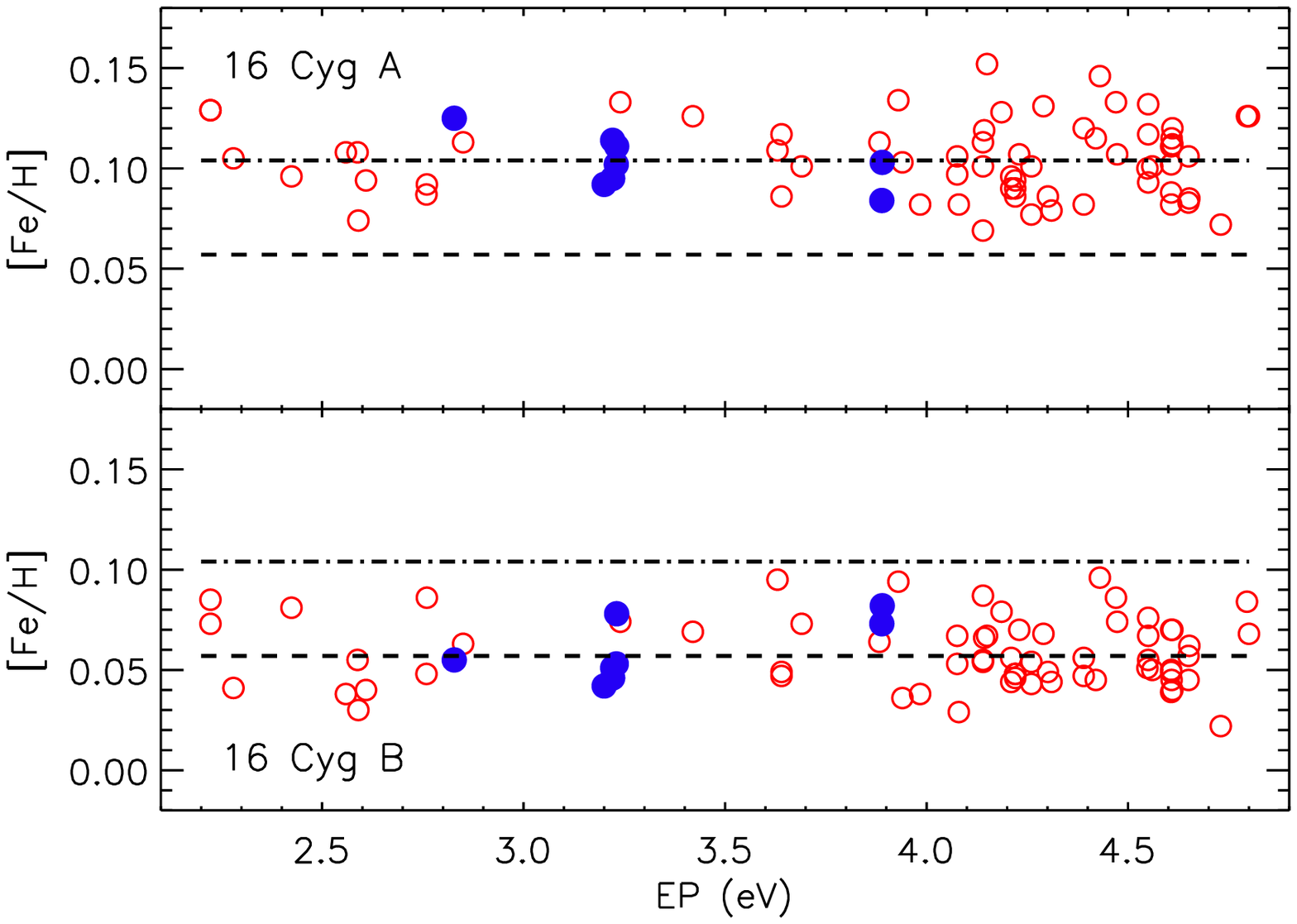}
\includegraphics[bb=60 360 560 712,width=8.7cm]{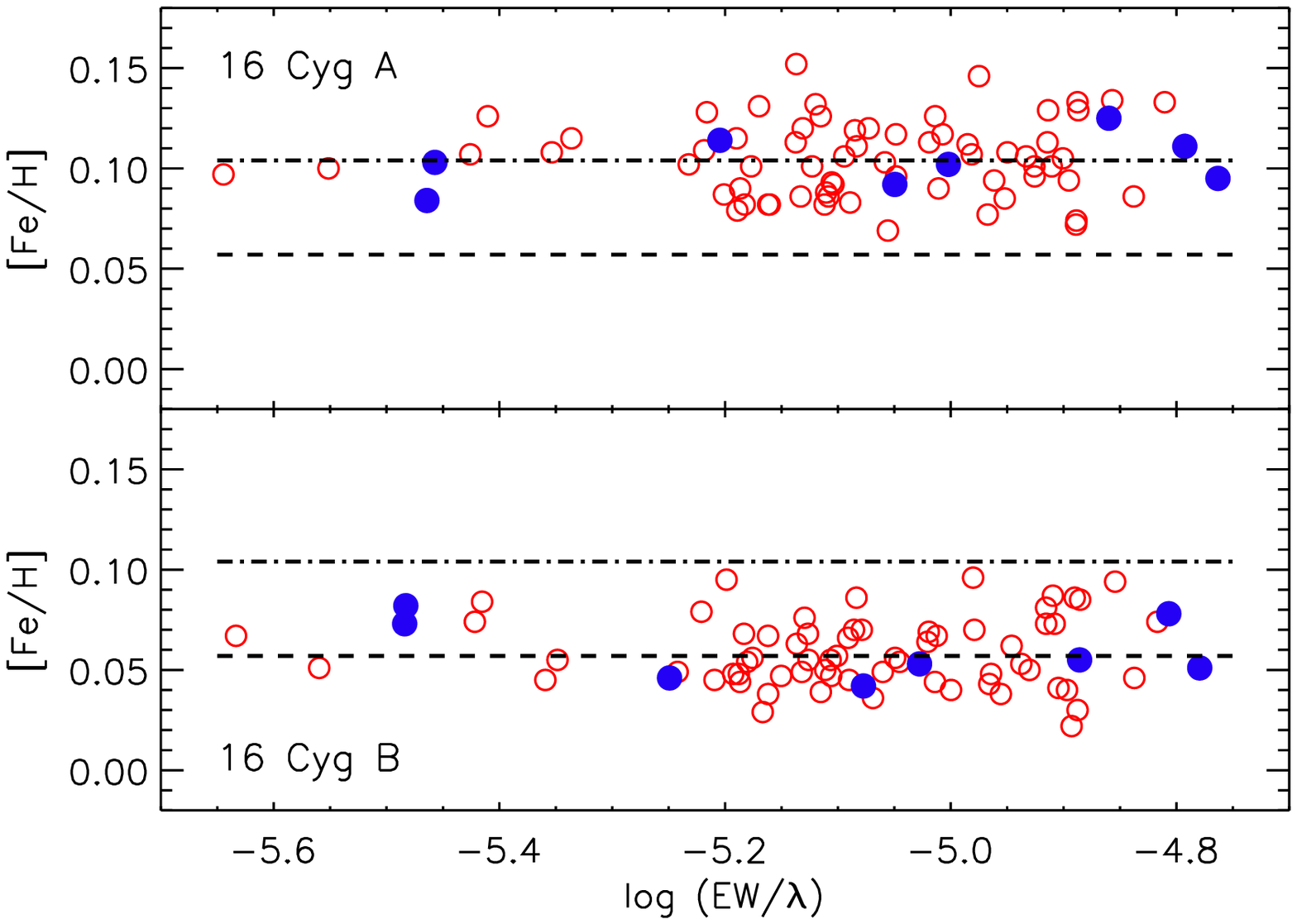}
\caption{Top panels: $\feh$ values determined on a line-by-line basis with respect to our solar spectrum as a function of excitation potential for \cyga\ and B; \fei\ (\feii) lines are shown with open (filled) circles. The dot-dashed line is at $\feh=+0.104$ while the dashed line is at $\feh=+0.057$, which are the mean $\feh$ values for \cyga\ and B, respectively. Bottom panels: as in the top panels but as a function of reduced equivalent width.}
\label{f:spec}
\end{figure}

Using the excitation/ionization equilibrium method we obtain: $\teff=5818\pm38$\,K, $\logg=4.31\pm0.05$, $\feh=+0.104\pm0.018$ for \cyga; and $\teff=5742\pm30$\,K, $\logg=4.31\pm0.05$, $\feh=+0.057\pm0.017$ for \cygb\ (see Fig.~\ref{f:spec}). If we use Kurucz no-overshoot models instead of MARCS we find a shift of only +5\,K in $\teff$, +0.01\,dex in $\logg$, and $+0.002$\,dex in $\feh$, systematically consistent for both stars. Moreover, adopting a slightly different solar microturbulence: $v_t=1.0\pm0.1$\,\kms, we find changes of only $\pm0.003$\,dex in $\feh$ and, as expected, shifts of $\pm0.1$\,\kms\ in the microturbulent velocities of \cyga\ and B. This shows that the choices of model atmosphere flavor and solar microturbulence have an almost negligible impact on our differential abundance results.

The calculation described in this section suggests that \cyga\ is slightly more metal-rich than \cygb. However, since there is a non-negligible difference in the effective temperature of the stars, small differential systematic effects might still be present and it is therefore important to check the parameters here derived using other independent methods.

\subsection{Photometric effective temperature}

\cite{casagrande10} have implemented the so-called infrared flux method (IRFM) to derive the effective temperatures of a large sample of dwarf and subgiant stars. The zero point of their temperature scale has been determined with an unprecedented precision of only 15\,K. Metallicity-dependent color-$\teff$ transformations provided in that work allow us to estimate $\teff$ values for the \cyga\ and B pair from their observed photometry. We used the $(B-V)$ and $(b-y)$ colors and our derived $\feh$ values to obtain: $\teff=5792$\,K for \cyga\ and $\teff=5733$\,K for \cygb. Photometric errors translate into an error of about 16\,K for both stars, similar to the color-to-color error. The uncertainties in $\feh$ translate into a 5\,K uncertainty in $\teff$. Adding these errors in quadrature we estimate a random error of 23\,K, which has to be added linearly to the systematic error in the zero point of the \cite{casagrande10} $\teff$ scale to estimate the uncertainties in our derived $\teff$ values. Thus, the effective temperatures we derived using colors have errors of 38\,K.

The color temperatures, which are based on the most reliable IRFM temperature scale to date, confirm that \cyga\ is warmer than \cygb, although the difference seems to be smaller when using the IRFM $\teff$ scale (59\,K) instead of the spectroscopic method described in the previous section (76\,K). Nevertheless, the $\teff$ values from the two methods are in good agreement within the 1$\sigma$ errors.

\subsection{Model fits to the H$\alpha$ line profile} \label{s:halpha}

The wings of the Balmer lines are known to be sensitive to $\teff$ and if $\logg$ and $\feh$ can be estimated with high precision independently, as in our case, model fits to the H$\alpha$ line profile can provide very precise $\teff$ values \cite[e.g.,][]{fuhrmann94,barklem02,ramirez06}. This is particularly the case if a good continuum normalization is made using very high quality spectra.

\begin{figure}
\includegraphics[bb=70 360 560 712,width=8.7cm]{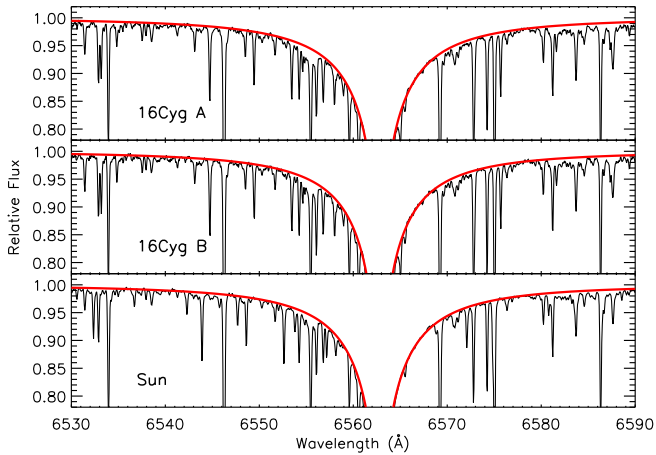}
\caption{Observed normalized H$\alpha$ line profiles in our \cyga, \cygb, and daysky (Sun) spectra. Best fit models for these lines are also shown.}
\label{f:halpha}
\end{figure}

We used least squares minimization to find the best model fits to our H$\alpha$ data as well as the formal errors in $\teff$. Careful inspection of our spectra allowed us to find several wavelength windows free from spectral lines, both stellar and telluric, other than H$\alpha$, which were then used in the model fits. The same windows were used for the three spectra. In Fig.~\ref{f:halpha} we show the observed normalized H$\alpha$ lines for \cyga, B, and the Sun (daysky) along with their best fit models. The problems described in Sect.~\ref{s:data} for the solar spectrum are not as important for this broad feature, in particular because we are only interested in the wings, so we can use reliably our daysky spectrum as solar reference in this case. For our solar data, we derive $\teff=5732\pm32$\,K, similar to the value obtained by \cite{barklem02}.\footnote{Our error bar here corresponds to observational noise only. \cite{barklem02} studied systematic errors very carefully and obtained an error of about 80\,K for their solar analysis. To remain consistent with the rest of our strictly differential work, we consider the observational errors only.} This suggests that there is still room for improvement in the modeling of Balmer lines. Hereafter, we apply a zero point correction of +45\,K to the effective temperatures derived from H$\alpha$ since this brings our derived solar $\teff$ up to the nominal value of 5777\,K.

We computed $\teff$ values from the H$\alpha$ line profile fits for a small grid of $\logg=4.2,4.3,4.4$ and $\feh=0.00,0.05,0.10$. Then, we used the grid of $\teff$ results to interpolate to our preferred $\logg$ and $\feh$ parameters as well as to compute the error introduced by uncertainties in $\logg$ and $\feh$. With our final $\logg$ and $\feh$ values we obtain: $\teff=5819\pm25$\,K for \cyga\ and $\teff=5762\pm26$\,K for \cygb. An error of 0.02\,dex in $\logg$ translates into only 2\,K whereas an error of 0.03\,dex in $\feh$ translates into about 1\,K of $\teff$. The uncertainties in our derived values of $\logg$ and $\feh$ thus have negligible impact on this calculation. Moreover, given the very high signal-to-noise ratios of our spectra in this region ($S/N\simeq500$, obtained because the spectrograph setup was chosen to maximize $S/N$ at 6562\,\AA), our $\teff$ values derived from H$\alpha$ line profile fitting are the most reliable, as the smaller errors in comparison with $\teff$ values derived with other methods suggest.

\bigskip

Our three estimates of $\teff$ are in excellent agreement with each other for both \cyga\ and B. Since these techniques are independent, we can obtain a final $\teff$ value by averaging them, using the estimated errors as weights. We find: $\teff=5813\pm18$\,K for \cyga\ and $\teff=5749\pm17$\,K for \cygb.\footnote{We calculate the weighted mean value ($x$) and error ($\sigma$) of $N$ data points $x_i$ with errors $\sigma_i$ as follows: $x=\sum(x_i/\sigma_i^2)/\sum(1/\sigma_i^2)$, $\sigma^2=1/\sum(1/\sigma_i^2)$. We note, however, that this formula for $\sigma$ does not take into account possible systematic errors and could also under- or over-estimate the error if the $\sigma_i$ values are not realistic. The error computed from the sample variance is in principle more reliable: $\sigma^2=\{\sum[(x_i-x)^2/\sigma_i^2)]/\sum(1/\sigma_i^2)\}/(N-1)$. Using the latter we find $\teff$ errors of 8 and 9\,K for \cyga\ and B, respectively, but we adopt the former, more conservative error bars.} The degree of precision with which we have been able to measure the effective temperatures of these stars is almost unprecedented for any other star but the Sun. The only other exception is the case of the solar twin star HIP\,56948, for which \cite{melendez11} report a $\teff$ with an error of 7\,K.

\subsection{Trigonometric surface gravity}

The parallax of the 16\,Cyg pair has been accurately measured by the Hipparcos mission \citep{perryman97}. The new reduction by \cite{vanleeuwen07} gives values of $47.44\pm0.27$\,mas for \cyga\ and $47.14\pm0.27$\,mas for \cygb. The $V$ magnitudes of these stars are also precisely known: $V(\mathrm{A})=5.959\pm0.009$, $V(\mathrm{B})=6.228\pm0.019$. Using these numbers we can infer the absolute $V$ magnitudes of the stars: $M_V(\mathrm{A})=4.340\pm0.015$, $M_V(\mathrm{B})=4.595\pm0.023$. Since their effective temperatures are also known with very high precision, we can reliably place the stars on the HR diagram and compare their location with theoretical stellar evolution predictions (isochrones) to infer their masses, ages, and surface gravities.

Our particular implementation of stellar parameter determination using isochrones is described in \citet[][see also \citealt{baumann10}]{melendez11}. Our method follows the probabilistic scheme explained in R03 (their Sect.~3.4) and also \citet[][their Appendix~A]{allende04:s4n} but the choice of isochrones is different. Basically, all isochrone points within a radius of 3$\sigma$ from the observed parameters are used to compute mass, age, and $\logg$ probability distribution functions, from which a most likely value and 68\,\% (1$\sigma$-like) and 95\%\ (2$\sigma$-like) confidence limits can be derived. Finely spaced Yonsei-Yale isochrones \cite[e.g.,][]{yi01,yi03,kim02} were used in these computations. 

Using our isochrone method we infer masses of $M(\mathrm{A})=1.05\pm0.02\,M_\odot$ and $M(\mathrm{B})=1.00\pm0.01\,M_\odot$. For the surface gravities we find: $\logg(\mathrm{A})=4.28\pm0.02$ and $\logg(\mathrm{B})=4.33\pm0.02$. Since the two stars were born at the same time, it is expected for the primary to be slightly more evolved given its somewhat larger mass.

\begin{figure}
\includegraphics[bb=80 370 560 710,width=8.7cm]{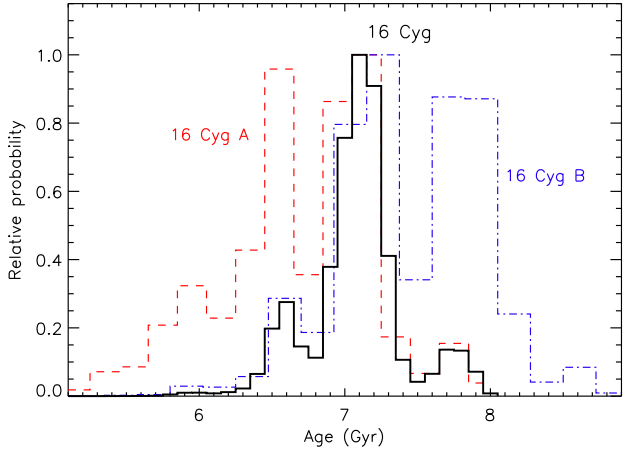}
\caption{Normalized age probability distributions for \cyga\ (dashed histogram), \cygb\ (dot-dashed histogram), and the 16\,Cyg binary system (solid line histogram).}
\label{f:age}
\end{figure}

The ages that we infer for the stars in the 16\,Cyg system are: $\tau(\mathrm{A})=7.15^{+0.04}_{-1.03}$\,Gyr and $\tau(\mathrm{B})=7.26^{+0.69}_{-0.33}$\,Gyr. Error bars correspond to the 1$\sigma$-like lower and upper limits. The age probability distributions (APDs) for these stars are shown in Fig.~\ref{f:age}. The asymmetry of these APDs explains the very different upper and lower 1$\sigma$ errors given above. Although they are somewhat broad, the APDs of \cyga\ and B peak at nearly the same age. A much more precise age can be obtained by combining the two APDs, as shown with the solid line histogram in Fig.~\ref{f:age}. This combined 16\,Cyg APD is simply the product of the APDs of the components of the 16\,Cyg binary system, normalized to 1 at the maximum. Using the combined APD we infer an age of $7.10^{+0.18}_{-0.35}$\,Gyr for 16\,Cyg. Note that this very small age error bar does not include systematic errors due to model uncertainties or errors introduced by statistical biases (see below). It represents only the internal precision of our method when applied to a binary system.

Stellar parameters derived from isochrones are subject to a number of systematic errors as well as statistical biases due to isochrone sampling which are described in detail by, for example, \citet[][their Sect.~4.5.4]{nordstrom04}. Bayesian approaches have been proposed to tackle these problems \cite[e.g.,][]{pont04,jorgensen05,dasilva06,casagrande11}. We can compare our results with the Bayesian implementation by \cite{dasilva06} since they provide a very convenient online tool that uses their method.\footnote{http://stev.oapd.inaf.it/cgi-bin/param} We find $\logg(\mathrm{A})=4.26\pm0.01$ and $\logg(\mathrm{B})=4.31\pm0.01$, in excellent agreement with our estimates even though the choice of isochrones is different (\citeauthor{dasilva06} use PADOVA isochrones; e.g., \citealt{bertelli94,girardi00}) and we do not correct for statistical biases. The masses derived with the \cite{dasilva06} online tool are about $0.02\,M_\odot$ lower but the difference between \cyga\ and B is nearly identical. For the stellar ages, \citeauthor{dasilva06} online tool suggests $6.7\pm0.5$\,Gyr (A) and $7.6\pm0.7$\,Gyr (B). These values are also consistent with our derived age for the 16\,Cyg system considering the 1$\sigma$ uncertainties.

Note that the excitation/ionization method of atmospheric parameter determination was not able to resolve the small difference in $\logg$ found in this section although the results are also consistent within 1$\sigma$. Clearly, for nearby stars with well determined effective temperatures, parallaxes, and good photometry, the isochrone method is superior.

\subsection{Surface gravity from the \mgib\ lines}

The wings of certain strong spectral features are mildly sensitive to the stellar surface gravity. A particularly important feature in this regard is the \mgib\ triplet at about 5180\,\AA. The effective temperature and magnesium abundance must be known with high precision in order for this line to be useful in the determination of $\logg$.

\begin{figure}
\includegraphics[bb=50 360 560 715,width=8.7cm]{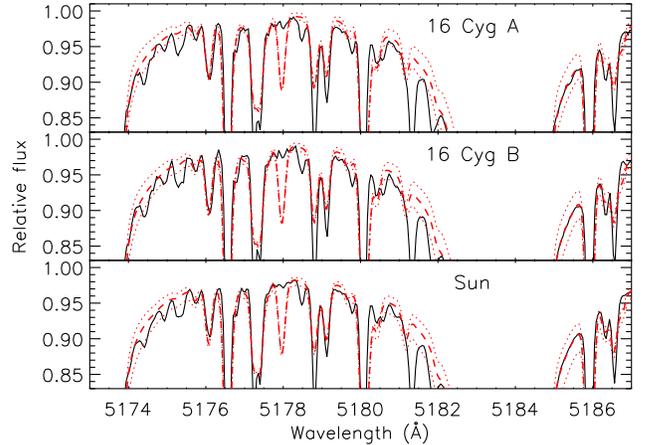}
\caption{Spectral region showing the strongest of the \mgib\ lines and the red wing of the middle \mgib\ line in \cyga, \cygb, and the Sun. Dashed lines show best fit models for each star (with their derived $\logg$ values) while dotted lines correspond to synthetic spectra with $\logg\pm0.2$.}
\label{f:mgib}
\end{figure}

Fig.~\ref{f:mgib} shows the spectral region around the strongest of the \mgib\ lines and the red wing of the middle one along with synthetic spectra computed for values of $\logg$ that reproduce best the regions free from lines other than the \mgib\ lines, such as those around 5174\,\AA\ and 5185\,\AA. We fine-tuned the transition probabilities of these lines to reproduce the solar \mgib\ lines with a Mg abundance of $A_\mathrm{Mg}=7.60$, as measured by \cite{asplund09:review}. For \cyga\ and B we used synthetic models with our final $\teff$ and $\feh$ values as well as our derived magnesium abundances. Synthetic profiles for $\logg\pm0.2$\,dex are also shown in Fig.~\ref{f:mgib} to illustrate the sensitivity of these features to $\logg$. From a least-squares minimization of the clean regions we find $\logg=4.27\pm0.04$ for \cyga\ and $\logg=4.33\pm0.04$ for \cygb, in excellent agreement with the values found in the previous section.

\bigskip

Similar to the case of $\teff$, the three methods used to derive $\logg$ are independent so a weighted average can be used to determine the best solution. We find: $\logg(A)=4.282\pm0.017$, $\logg(B)=4.328\pm0.017$. These and other fundamental parameters derived in this section are summarized in Tables~\ref{t:tefflogg} and \ref{t:parameters}.

\begin{table}
\caption{Effective temperature and surface gravity determinations}
\label{t:tefflogg}
\centering
\begin{tabular}{ccc}\hline\hline
$\teff$ (K) & \cyga & \cygb \\ \hline
Fe lines & $5818\pm38$ & $5742\pm30$ \\
Colors (IRFM) & $5792\pm38$ & $5733\pm38$ \\
H$\alpha$ & $5819\pm25$ & $5762\pm26$ \\ \hline
Adopted & $5813\pm18$ & $5749\pm17$ \\ \hline\hline
$\logg$ [cgs] & \cyga & \cygb \\ \hline
Fe lines & $4.31\pm0.05$ & $4.31\pm0.05$ \\
Trigonometric & $4.28\pm0.02$ & $4.33\pm0.02$ \\
\mgib & $4.27\pm0.04$ & $4.33\pm0.04$ \\ \hline
Adopted & $4.282\pm0.017$ & $4.328\pm0.017$ \\ \hline
\end{tabular}
\end{table}

\begin{table}
\caption{Basic data and derived parameters for the 16\,Cyg binary system}
\label{t:parameters}
\centering
\begin{tabular}{c||cc}\hline\hline
& \cyga & \cygb \\ \hline
$V$ (mag)   & $5.959\pm0.009$ & $6.228\pm0.019$ \\
$\pi$ (mas) &  $47.44\pm0.27$ & $47.14\pm0.27$  \\
$M_V$ (mag) & $4.340\pm0.015$ & $4.595\pm0.023$ \\
$\teff$ (K) & $5813\pm18$ & $5749\pm17$ \\
$\logg$ [cgs] & $4.282\pm0.017$ & $4.328\pm0.017$ \\
Mass ($M_\odot$) & $1.05\pm0.02$ & $1.00\pm0.01$ \\
\hline
\end{tabular}
\end{table}

\section{Elemental abundances}

Curve of growth analysis within MOOG was used to determine elemental abundances from equivalent width measurements of lines in the spectra of \cyga, B, and the Sun. We measured 288 lines for 22 elements; their $EW$s are given in Table~\ref{t:lines}. Since the oxygen abundances inferred from the infrared triplet lines, which are the only oxygen lines we use, are severely affected by non-LTE effects \cite[e.g.,][]{kiselman93,fabbian09}, and reliable non-LTE computations can be found in the literature, we used the non-LTE correction tables by \cite{ramirez07} to take this systematic effect into account. Lines from two species (neutral and singly ionized) of Sc, Ti, Cr, and Fe were available; for the other elements only one species was measured. Good agreement was found for the abundances derived from two species of the same element except for Sc (see below). The abundances of Nd and Eu were determined using spectrum synthesis of a few carefully selected features. Spectrum synthesis was also employed to measure lithium abundances from the 6708\,\AA\ Li doublet.

\begin{figure}
\includegraphics[bb=40 360 450 690,width=8.7cm]{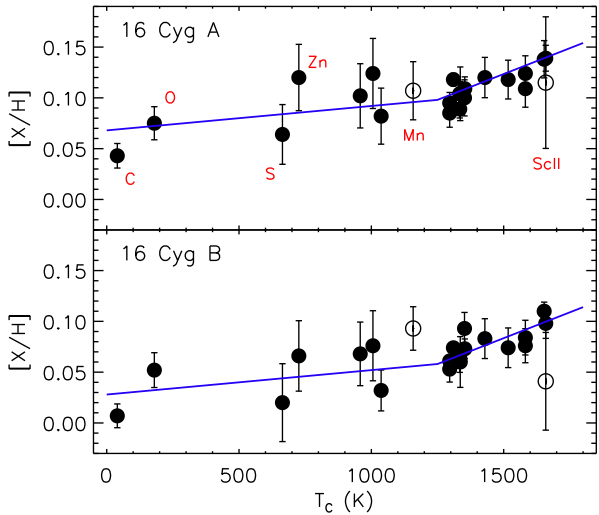}
\caption{Elemental abundances of \cyga\ and B relative to our measured solar abundances as a function of dust condensation temperature for light elements $(Z\leq30)$. Solid lines show the mean trend of solar twin stars according to \cite{melendez09:twins} but shifted upwards in each case to match the mean abundance of volatile ($\tc<900$\,K) elements. Open circles correspond to Mn, Co, and Sc\,\textsc{ii}.}
\label{f:trends_solar}
\end{figure}

\begin{figure}
\includegraphics[bb=40 360 450 690,width=8.7cm]{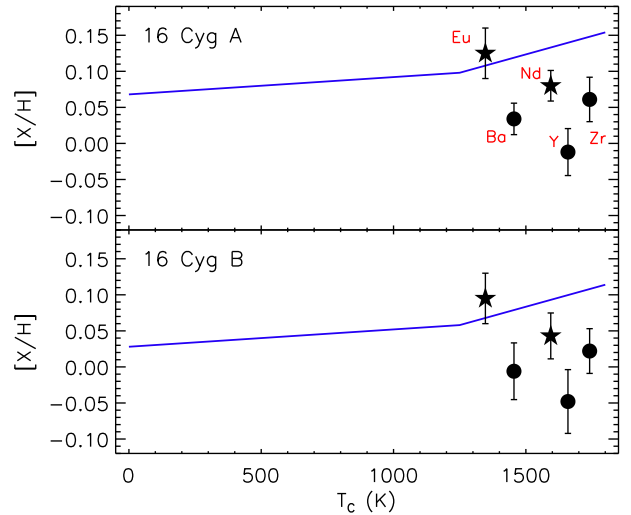}
\caption{As in Fig.~\ref{f:trends_solar} for the heavy elements $(Z>30)$. Stars (circles) denote elemental abundances measured with spectrum synthesis (equivalent widths).}
\label{f:trends_solar_heavy}
\end{figure}

We measured differential abundances on a line-by-line basis, which reduces the impact of errors in the atomic data and systematics in the modeling, given that these are all similar stars. Abundance ratios relative to solar ([X/H]) are listed in Table~\ref{t:abund}. Errors in this table correspond to the standard error of the line-to-line scatter added in quadrature with the random error introduced by the uncertainties in our derived atmospheric parameters. Fig.~\ref{f:trends_solar} shows the [X/H] abundances as a function of the dust condensation temperature sequence by \cite{lodders03} for the light elements ($Z\leq30$). Heavy element ($Z>30$) abundances are shown separately in Fig.~\ref{f:trends_solar_heavy}. In both Figs.~\ref{f:trends_solar} and \ref{f:trends_solar_heavy} we plot with a solid line the mean trend of solar twin stars by M09, shifted to match approximately the abundance pattern of $Z\leq30$ elements for each star.

\begin{table*}
\caption{Elemental abundances relative to solar ([X/H]) and differential between \cyga\ and B ($\Delta\mathrm{[X/H]}$).}
\label{t:abund}
\centering
\begin{tabular}{rcccccccc}\hline\hline
 & Species & $\tc$ (K) & [X/H] & error & [X/H] & error & $\Delta\mathrm{[X/H]}$ & error \\
 & & & \multicolumn{2}{c}{(\cyga)} & \multicolumn{2}{c}{(\cygb)} & \multicolumn{2}{c}{(\cyga--\cygb)} \\ \hline
 6.0 &    C &   40 &  0.043 &  0.012 &  0.007 &  0.012 &  0.037 &  0.016 \\
 8.0 &    O &  180 &  0.075 &  0.016 &  0.052 &  0.018 &  0.024 &  0.024 \\
11.0 &   Na &  958 &  0.102 &  0.032 &  0.068 &  0.031 &  0.034 &  0.020 \\
12.0 &   Mg & 1336 &  0.104 &  0.026 &  0.060 &  0.026 &  0.043 &  0.013 \\
13.0 &   Al & 1653 &  0.138 &  0.018 &  0.110 &  0.018 &  0.028 &  0.023 \\
14.0 &   Si & 1310 &  0.118 &  0.006 &  0.074 &  0.004 &  0.044 &  0.008 \\
16.0 &    S &  664 &  0.064 &  0.029 &  0.020 &  0.030 &  0.044 &  0.016 \\
19.0 &    K & 1006 &  0.124 &  0.034 &  0.076 &  0.034 &  0.048 &  0.028 \\
20.0 &   Ca & 1517 &  0.118 &  0.019 &  0.074 &  0.018 &  0.044 &  0.017 \\
21.0 &  ScI & 1659 &  0.115 &  0.065 &  0.041 &  0.065 &  0.074 &  0.027 \\
21.1 & ScII & 1659 &  0.139 &  0.013 &  0.098 &  0.012 &  0.041 &  0.011 \\
22.0 &  TiI & 1582 &  0.109 &  0.018 &  0.076 &  0.016 &  0.033 &  0.023 \\
22.1 & TiII & 1582 &  0.124 &  0.017 &  0.084 &  0.017 &  0.040 &  0.009 \\
23.0 &    V & 1429 &  0.120 &  0.020 &  0.083 &  0.020 &  0.037 &  0.026 \\
24.0 &  CrI & 1296 &  0.085 &  0.014 &  0.053 &  0.013 &  0.033 &  0.018 \\
24.1 & CrII & 1296 &  0.095 &  0.010 &  0.061 &  0.011 &  0.034 &  0.013 \\
25.0 &   Mn & 1158 &  0.107 &  0.029 &  0.093 &  0.028 &  0.014 &  0.022 \\
26.0 &  FeI & 1334 &  0.104 &  0.012 &  0.061 &  0.011 &  0.042 &  0.016 \\
26.1 & FeII & 1334 &  0.089 &  0.009 &  0.064 &  0.010 &  0.025 &  0.013 \\
27.0 &   Co & 1352 &  0.100 &  0.018 &  0.093 &  0.016 &  0.007 &  0.018 \\
28.0 &   Ni & 1353 &  0.109 &  0.012 &  0.073 &  0.010 &  0.036 &  0.014 \\
29.0 &   Cu & 1037 &  0.082 &  0.028 &  0.032 &  0.028 &  0.051 &  0.019 \\
30.0 &   Zn &  726 &  0.120 &  0.033 &  0.066 &  0.033 &  0.053 &  0.008 \\
39.1 &    Y & 1659 & -0.012 &  0.033 & -0.048 &  0.044 &  0.036 &  0.016 \\
40.1 &   Zr & 1741 &  0.061 &  0.031 &  0.022 &  0.031 &  0.039 &  0.018 \\
56.1 &   Ba & 1455 &  0.034 &  0.022 & -0.006 &  0.039 &  0.040 &  0.020 \\
60.1 &   Nd & 1594 &  0.080 &  0.021 &  0.043 &  0.032 &  0.030 &  0.035 \\
63.1 &   Eu & 1347 &  0.125 &  0.035 &  0.095 &  0.035 &  0.037 &  0.023 \\

\hline
\end{tabular}
\end{table*}

The weighted mean [X/H] abundances from all species studied, which we denote by [M/H], are $+0.103\pm0.023$ for \cyga\ and $+0.069\pm0.026$ for \cygb. The 16\,Cyg binary system is therefore more metal-rich than the Sun but, interestingly, a small difference in the overall metal content appears to make \cyga\ more metal-rich than \cygb.

Considering only the light elements ($Z\leq30$, Fig.~\ref{f:trends_solar}), the observed abundances of \cyga\ and B follow the typical pattern of solar twin stars, i.e., they correlate with dust condensation temperature as in M09. Note that the latter was obtained from the average abundances of 11 solar twin stars, resulting in a very small abundance error (of order 0.01\,dex). However, for individual stars, the dispersion seen in Fig.~\ref{f:trends_solar} is fully consistent with the analysis by M09.

On the other hand, the heavy element abundances ($Z>30$, Fig.~\ref{f:trends_solar_heavy}) do not follow the typical solar twin behavior. Strictly speaking, \cyga\ and B are solar analogs and not solar twins, mainly due to their slightly lower surface gravities, which makes them more evolved than the Sun.  Compared to the Sun, the 16\,Cyg system is about 2.5\,Gyr older and it is therefore likely that the anomalous abundances of the heavy elements (when compared to solar twins) are due to Galactic chemical evolution effects. We discuss these in detail in Sect.~\ref{s:gce}.

In principle, the mean metallicities ([M/H]) of \cyga\ and B are consistent within 1$\sigma$. We should point out, however, that this analysis uses our solar spectrum as reference. Although obtained with the same instrument, which is ideal, our solar spectrum was not observed in the exact same way the stellar spectra were taken. Important differences in the data reduction procedures are also present. Furthermore, our reference spectrum is based on observations of the daysky, which is known not to be an exact replica of the solar spectrum, although we attempted to minimize its impact by using daysky spectra at small angular distances from the Sun, as explained in Sect.~\ref{s:data}.

\begin{figure}
\includegraphics[bb=70 360 450 800,width=8.7cm]{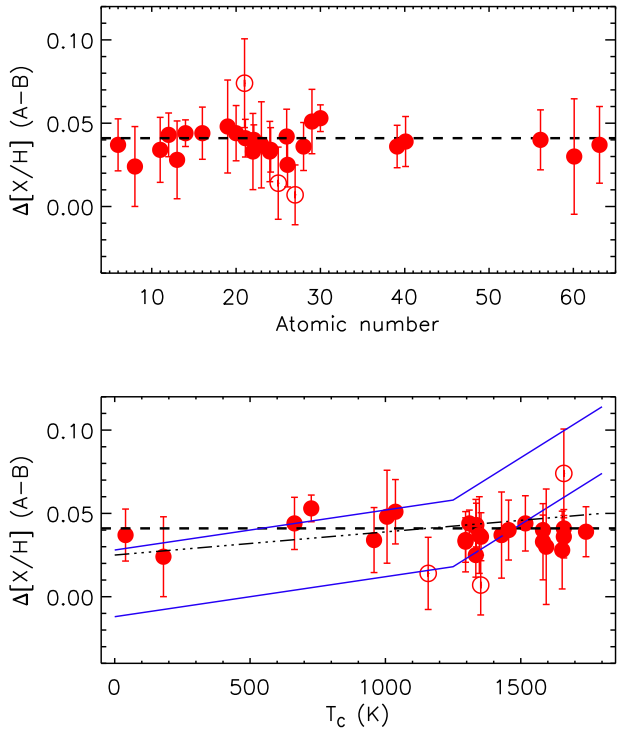}
\caption{Top panel: elemental abundance difference between \cyga\ and B as a function of atomic number. Open symbols show the three species more discrepant from the mean: Sc\,\textsc{i} (21), Mn (25), and Co (27). Bottom panel: as in the top panel for the abundance differences versus dust condensation temperature. The dashed line is at +0.041 while the solid lines represent the mean trend of solar twins by \cite{melendez09:twins} with two arbitrary offsets. The dot-dashed line corresponds to a slope of $1.4\times10^{-5}$\,dex\,K$^{-1}$, as derived by \cite{laws01}.}
\label{f:trends}
\end{figure}

Inspection of Figs.~\ref{f:trends_solar} and \ref{f:trends_solar_heavy} shows clearly that the element to element scatter is very similar in \cyga\ and B. This suggests that the difference between A and B might be better established by comparing them directly rather than using our possibly faulty solar spectrum. Galactic chemical evolution effects will also be minimized, if not eliminated, with this approach. Thus, we measured differential abundances of \cyga, again on a line-by-line basis, but using \cygb\ as the reference star. We denote these differential abundances with the symbol $\Delta\mathrm{[X/H]}$. The results are given in Table~\ref{t:abund}. Similar to the errors in [X/H], the errors listed in Table~\ref{t:abund} for $\Delta\mathrm{[X/H]}$ correspond to the standard error of the line-to-line scatter added in quadrature with the errors introduced by uncertainties in our derived stellar parameters. Fig.~\ref{f:trends} shows the derived \cyga\ -- \cygb\ relative abundances as a function of atomic number and dust condensation temperature. Note that in this case both the light and heavy elements behave similarly.

Fig.~\ref{f:trends} shows that not only are the 1$\sigma$ errors of the A--B relative abundances small but it also reveals that the element-to-element dispersion is extremely small. A weighted average of all species measured suggests a metallicity offset of $\Delta\mathrm{[M/H]}=+0.040\pm0.010$\,dex for \cyga\ relative to \cygb. The three species that depart the most from this average value are Sc\,\textsc{i}, Mn, and Co. It is tempting to attribute these discrepant data points to strong differential non-LTE effects (see \citealt{zhang08} for Sc\,\textsc{i}, \citealt{bergemann07} for Mn, and \citealt{bergemann08} for Co; see also \citealt{bergemann10}) since there is a difference of about 65\,K in $\teff$ between the two stars in addition to a difference of 0.05\,dex in $\logg$. Furthermore, we note that the Sc\,\textsc{ii} abundance, which is known to be more reliable than Sc\,\textsc{i}, is consistent with the derived $\Delta\mathrm{[M/H]}$ while the 1$\sigma$ line-to-line scatter (not the standard error) of the Mn abundance difference is the largest of all species analyzed with equivalent widths. A more proper treatment of Co line formation requires consideration of hyperfine structure (HFS) splitting, which we tested for three Co\,\textsc{i} features using HFS constants by \cite{kurucz95}. In this case we obtained $\Delta\mathrm{[Co/H]}=+0.007\pm0.018$, in excellent agreement with the value derived from equivalent widths and ignoring HFS. Perhaps a combination of HFS and non-LTE effects could explain our Co abundances. \cite{bergemann10}, for example, calculate a non-LTE correction of 0.13\,dex for the Sun, which is relatively large. Excluding arbitrarily these three discrepant data points (Sc\,\textsc{i}, Mn, and Co), which are likely affected by small but non-negligible systematic errors in high precision work, the metallicity offset is: $\Delta\mathrm{[M/H]}=+0.041\pm0.007$ (A--B). Hereafter, we discuss our results ignoring the Sc\,\textsc{i}, Mn, and Co abundances.

Note that the 1$\sigma$ uncertainty of the metallicity difference that we inferred above corresponds to about 1\,\% on a linear scale. Interestingly, the 1$\sigma$ error of our measured $EW$s is also about 1\,\%. This suggests that the error in our very precise elemental abundance determinations, which are based on curve of growth analysis (exclusively on the linear part) with the $EW$s as input data, is dominated by the observational errors (elements Nd and Eu are the only exceptions, since their abundances were measured using line profile fitting). Thus, it should be in principle possible to improve upon these estimates using even higher quality data.

We have explored the impact of systematic errors in our derived atmospheric parameters by arbitrarily modifying them and recomputing the abundances. First, we tried increasing and decreasing the $\teff$ values of both stars by 100\,K. Interestingly, in both cases we find that the A--B metallicity difference decreases to about 0.03\,dex but the element-to-element scatter increases to 0.008\,dex. Adopting identical $\logg=4.30$ values for both stars increases the metallicity difference to +0.046\,dex but in this case the element-to-element scatter increases significantly to 0.013\,dex. The fact that the element-to-element scatter naturally minimizes for our derived parameters suggests that they are very reliable.

We have also re-determined our abundances using the differential parameters derived by \cite{takeda05}, who claims that there is no metallicity difference between \cyga\ and B. We kept our \cygb\ parameters fixed and adopted \cyga\ parameters consistent with \citeauthor{takeda05}'s. The inferred metallicity difference in this case is $+0.032\pm0.013$, i.e., we find that \cyga\ is still more metal-rich than \cygb. However, for these parameters we detect a small negative trend for the abundance differences and condensation temperature which makes the difference smaller for refractory elements like iron. For those elements the metallicity difference is about $+0.02\pm0.01$\,dex. It seems possible that a combination of these effects is responsible for the \cite{takeda05} results but we admit that we are unable to reproduce exactly his derived elemental abundances.

Our main result that \cyga\ is more metal-rich than \cygb\ was already suggested in the high precision work by \cite{laws01}. They found, however, a metallicity difference of $+0.025\pm0.009$ based only on the analysis of iron lines. As a note added in proof they commented that abundances of 13 other elements were measured and showed a small positive correlation with dust condensation temperature. Their derived slope is shown in the bottom panel of our Fig.~\ref{f:trends}. Depending on the elements that one uses to calculate the slope, it could be possible to find good agreement between our data and the \citeauthor{laws01} result. Nevertheless, using our abundances we obtain essentially a zero slope ($-0.2\pm0.6\times10^5$\,dex\,K$^{-1}$), which is inconsistent with the slope derived by \citeauthor{laws01} ($+1.4\pm0.5\times10^5$\,dex\,K$^{-1}$). We argue that our results are more robust owing to the larger number of elements analyzed and the very high precision achieved for the overall metallicity difference.

Finally, using Kurucz' model atmospheres we find $\Delta\mathrm{[M/H]}=+0.039\pm0.007$ (no-overshoot models) and $\Delta\mathrm{[M/H]}=+0.040\pm0.007$ (overshoot models). Thus, we demonstrate again that the choice of model atmosphere grid has an almost negligible impact on our results.

\section{Discussion} \label{s:discussion}

\subsection{Planet signatures in stellar abundances}

High precision elemental abundance work in solar twins and analogs has revealed a possible connection between the properties of a star and the process of planet formation that occurred around it. The work by M09 showed that, compared to the majority ($\simeq$85\,\%) of solar twin stars, the Sun exhibits a deficiency of refractory elements relative to volatiles (of order 20\,\%). They proposed that this solar abundance anomaly could be due to the formation of the terrestrial planets in the solar system since these objects are rich in refractory elements. Thus, the missing mass of refractory elements in the solar photosphere would be currently located in the terrestrial planets. This hypothesis assumes that the majority of solar twin stars did not form as many rocky objects as in the solar system or that if they did, they were accreted by the star at later times, possibly through orbital migration. Only about 15\,\% of solar twin stars show a photospheric composition indistinguishable from the solar one. These objects are potentially terrestrial planet hosts. Later work by R09 and \cite{ramirez10} confirmed the observational results of M09 and expanded the sample towards higher metallicities, where they found that, based on the abundances measured, the fraction of terrestrial planet hosts increases towards higher [Fe/H].

Thus, it seems likely that a ``planet signature'' has been imprinted in the chemical composition of the host stars and, arguably, it appears to relate only to the formation of rocky objects. Indeed, M09 showed that metal-rich solar analogs with giant planets detected have abundance patterns relative to iron similar to those of the majority of solar twins, suggesting that the solar abundance anomaly could not be related to the formation of the gas giants. This type of analysis, however, should be examined more carefully because the impact of systematic errors and Galactic chemical evolution could be important in this higher $\feh$ regime \cite[see also][]{schuler11}. Data by independent groups have confirmed the observational result by M09 and R09 \citep{ramirez10,gonzalez-hernandez10,gonzalez10}. Although the planet formation signature interpretation has been contended by \cite{gonzalez-hernandez10}, a response to their criticism has been provided by \cite{ramirez10}. Alternative scenarios to explain the solar abundance anomaly have also been explored, without success \citep{ramirez10,melendez11,kiselman11}.

The planet signature hypothesis described above has been quantitatively tested by \cite{chambers10}, who computed the effect of adding two Earth masses ($2M_\oplus$) of Earth-like material and $2M_\oplus$ of CM chondrite material to the present-day solar convective envelope. He finds that in this experiment the solar abundance anomaly would disappear, i.e., the Sun would not show the deficiency of refractory elements anymore. Thus, his calculations support the hypothesis that the missing mass of rock in the solar photosphere is currently locked-up in the rocky bodies of the solar system. A similar estimate has been recently made by \cite{melendez11} who suggest an even larger amount of CM chondrite material ($4M_\oplus$). Although such large mass is not observed in the present-day asteroid belt, it has been suggested that small rocky bodies were very abundant when the Sun was formed, but most of the material from the asteroid belt was probably removed by strong resonances \cite[e.g.,][]{weidenschilling77,chambers01,petit01}.

Excluding the heavy elements (i.e., for $Z\leq30$), the abundance patterns of both \cyga\ and \cygb\ are compatible with those observed in most solar twin stars (Fig.~\ref{f:trends_solar}). Within the planet signature scenario, this implies that none of these objects host terrestrial planets or that any rocky body that once formed there has been accreted by the host star. Perhaps binary systems like 16\,Cyg are not stable enough to retain any terrestrial planets that might form. The gas giant planet around \cygb\ is highly eccentric, a property that has been attributed by some authors to the influence of the stellar companion \citep[e.g.,][]{cochran97,mazeh97}. The lack of rocky planets around \cygb\ could also be related to the presence of the gas giant planet. As evidenced by the Kepler mission data, multiple systems with small planets rarely contain hot Jupiters \citep{latham11}. In any case, the abundances of \cyga\ and B are not unusual considering that only about 15\,\% of solar twins share the exact same (solar) composition.

\subsection{Impact of Galactic chemical evolution} \label{s:gce}

The heavy elements clearly do not follow the typical solar twin trend (Fig.~\ref{f:trends_solar_heavy}). With the exception of Eu, the other $Z>30$ elements are found significantly below the mean solar twin trend. The Nd abundances are marginally consistent with the expected trend but the Y, Zr, and Ba abundances are irreconcilable with that pattern. These elements are produced primarily by neutron-capture reactions on timescales slow (the $s$-process) or rapid (the $r$-process) relative to the average $\beta$-decay rates of radioactive isotopes along the reaction chain. The $s$-process is associated with low- and intermediate-mass stars (1.3~$\lesssim M \lesssim$~8.0~$M_{\odot}$) on the asymptotic giant branch. The site (or sites) of the $r$-process is not known but is suspected to be associated with core collapse supernovae ($M \gtrsim$~8.0~$M_{\odot}$). These stellar timescales imply that $r$-process enrichment occurs earlier than $s$-process enrichment, which is confirmed by the observed heavy element abundance patterns in low metallicity stars in the Galactic halo and disk \cite[e.g.,][]{mcwilliam97}. 

Each of the $r$- and $s$-process contribute about half of the elements with $Z>30$ in the Sun, and, following pioneering work by \cite{cameron73}, the solar abundance of each isotope (or element) can be decomposed into $r$- and $s$-process fractions. For example, 72\% of the solar Y was produced via the $s$-process \cite[e.g.,][]{simmerer04}. Similarly, 81\% of Zr, 85\% of Ba, 58\% of Nd, and 3\% of Eu in the Sun were produced by the $s$-process. These fractions represent the chemical inventory of one point in the Galactic interstellar medium 4.5\,Gyr ago. The stars in the 16\,Cyg system formed 2.5~Gyr earlier and they are slightly deficient in $s$-process material relative to the Sun. As can be seen in Fig.~\ref{f:trends_solar_heavy}, elements with a significant $s$-process component, such as Y, Zr, and Ba, are the most deficient. Nd, which has roughly equal contributions from the $r$- and $s$-processes, is moderately deficient, and Eu, which has a minimal $s$-process contribution, is not deficient at all. The fact that the heavy elements in both \cyga\ and B show the same abundance pattern as a function of $\tc$ suggests that this is a primordial effect---the result of Galactic chemical evolution (GCE). The slightly super-solar metallicity of the system is not in contradiction, as the slope of [Ba/H] versus time is steeper than that for [Fe/H] \cite[e.g.,][]{edvardsson93}.

The works by M09 and R09 concentrated only on solar twin stars, therefore minimizing these relatively small GCE effects. Nevertheless, it is interesting to note that the solar twin abundances of Y, Zr, and Ba by R09 show larger star-to-star error than those of lighter elements (see their Figs.~1 and 3), a result that could be associated to the age span of their sample. The opposite effect of that observed in the 16\,Cyg system (i.e., neutron capture element abundances higher than the mean solar twin trend) has been recently observed in the younger solar twin star 18\,Sco (Mel\'endez et~al., in preparation), reinforcing the idea that this is related to GCE. Thus, our results for the heavy elements tell us that extreme caution must be practiced when interpretations of abundance ratios versus condensation temperature are made for stars that are not so similar to the Sun, in particular stars that are significantly older or younger, because GCE effects are not entirely negligible.

\subsection{Did \cyga\ accrete planetary material? Clues from lithium abundances}

\begin{figure}
\includegraphics[bb=75 360 560 690,width=8.7cm]{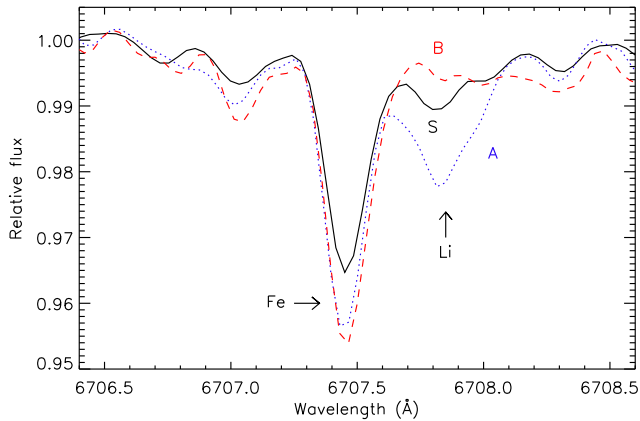}
\caption{Spectral region around the Li doublet feature (6708\,\AA). Our solar spectrum (solid line) is shown along spectra of \cyga\ (dotted line) and \cygb\ (dashed line).}
\label{f:lithium}
\end{figure}

Based on the discussion given in the previous sections, we are led to conclude that the metallicity difference between \cyga\ and B is not primordial but acquired. Either the photosphere of \cyga\ became more metal-rich or that of \cygb\ became more metal-poor after they were formed (or while they were being formed) from the same molecular cloud with the same initial chemical composition. \cite{laws01} suggested that \cyga\ accreted either 2.5 Earth masses of chondritic material or $0.3\,M_\mathrm{J}$ of gas giant material to explain its higher $\feh$ (they derived $\Delta\feh=+0.025\pm0.009$). They also argued that the late accretion of planetary material could increase the lithium abundance on the star's convective envelope. Indeed, inspection of the 6708\,\AA\ region clearly shows that \cygb\ is lithium poor compared to \cyga\ (Fig.~\ref{f:lithium}). Our derived Li abundances are: $A(\mathrm{Li})=1.34\pm0.04$ for \cyga\ and $A(\mathrm{Li})=0.73\pm0.06$ for \cygb, in good agreement with previously published estimates \citep{king97,luck06,takeda07,israelian09,baumann10}. Thus, \cyga\ has about 4.1 times ($\simeq0.6$\,dex) more lithium than \cygb\ in its photosphere.

\begin{figure}
\includegraphics[bb=30 10 400 387,width=9cm]{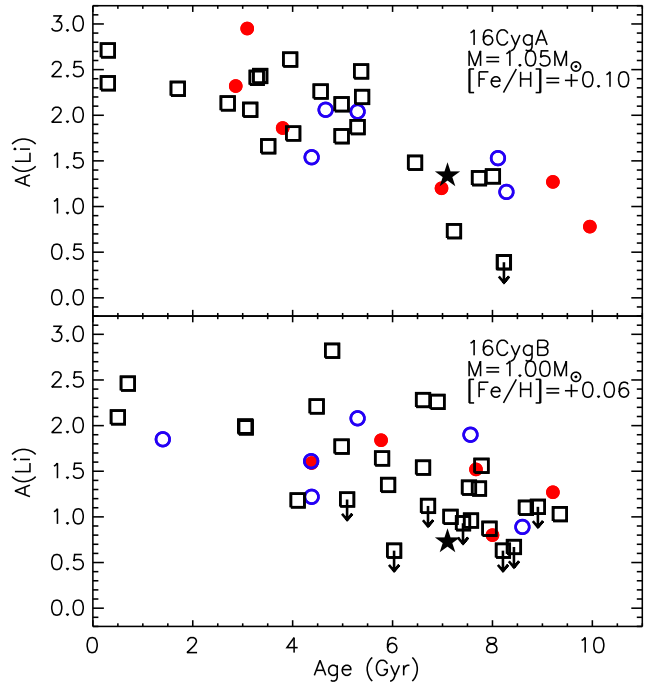}
\caption{Evolution of lithium abundance for stars of mass and metallicity similar to those of \cyga\ (top panel) and \cygb\ (bottom panel). In each panel the stellar mass is within $0.03\,M_\odot$ and metallicity within 0.05\,dex from those of the 16\,Cyg stars. Only stars with $\mathrm{age/error}>3$ are plotted. Upper limits on Li abundances are shown with downwards arrows. Open (filled) circles represent stars without (with) gas giant planets detected while open squares correspond to stars for which planet information is not available. Five-pointed stars show the location of \cyga\ (top panel) and \cygb\ (bottom panel).}
\label{f:liage}
\end{figure}

The behavior of Li abundances in solar-type stars is very complex and not yet fully understood \cite[e.g.,][]{chaboyer98,pinsonneault10}. In general, however, inspection of large samples clearly shows that at around 1\,$M_\odot$ and $\teff=5780$\,K a very steep slope of increasing Li abundances with stellar mass (or $\teff$) is present \cite[e.g.,][]{takeda07,baumann10}. Fitting a straight line to the Li abundance versus $\teff$ relation of stars in the 5700--5900\,K range we find that the lithium abundances of stars with $\teff$ values similar to those of \cyga\ and B should differ by a factor of 3.5 ($\simeq0.55$\,dex), i.e., very close to the observed value. Fish et~al.\ (in preparation) have recently derived lithium abundances of more than 700 solar-type stars and compiled and homogenized lithium abundance data for more than 1400 FGK stars. They also derive masses and ages in a consistent manner employing the exact same method used in this work. The lithium versus age trends for stars of mass and metallicity similar to those of \cyga\ and B, taken from the \citeauthor{fish11}\ catalog, are shown in Fig.~\ref{f:liage}. The lithium abundances of both \cyga\ and B are perfectly normal for stars of their mass, age, and metallicity, despite the somewhat large scatter, in particular for stars like \cygb.

If the convective envelope of \cyga, after it reached its present size, accreted 0.3\,$M_\mathrm{J}$ of planetary material that has not experienced lithium depletion, as suggested by \cite{laws01}, the photospheric lithium abundance would have increased by about 0.4\,dex (a factor of 2.5). Thus, if \cyga\ was polluted by planetary material, considering the mass effect on lithium depletion, the lithium abundance of \cyga\ should be about one order of magnitude larger than that of \cygb, which is significantly larger than observed. Such large lithium abundance for \cyga\ would also make the star somewhat lithium rich compared to stars of similar mass and metallicity (top panel in Fig.~\ref{f:liage}). Already the mass effect explains most, if not all, of the observed difference. The work by \cite{deliyannis00} on beryllium abundances also provides no evidence for accretion of planetary material by \cyga, although the authors warm that the errors are too large to make a firm conclusion.

\subsection{A signature of gas giant planet formation?}

A different possibility to explain the elemental abundance differences in the 16\,Cyg system is that it was produced by the formation of the planet around \cygb. Similar to M09, we propose that the convective zone of \cygb\ has been depleted of metals due to the formation of a planet, i.e., that the missing mass of metals in \cygb\ is currently locked-up in its planet.

Before exploring this hypothesis in detail, we should re-address the problem of lithium abundance. A number of authors have suggested that lower lithium abundances could be correlated with the presence of gas giants \cite[e.g.,][]{israelian09,gonzalez10:lithium}, although their preferred explanation is that it relates to physical processes that affect the rotational evolution of stars with planets and hence their lithium depletion history. A few planet-hosts and single stars are plotted in Fig.~\ref{f:liage}. Age, mass, and metallicity effects fully explain the observed abundances of both types of stars, without invoking the presence or absence of gas giant planets, consistent with the works by, for example, \cite{baumann10} and \cite{ghezzi10}. This is the case also for the 16\,Cyg stars; i.e., the significantly lower lithium abundance of \cygb\ ($\simeq0.6$\,dex relative to \cyga) is not due to the presence of its planet. In our hypothesis, however, a very small fraction of this difference ($\simeq0.04$\,dex) could be due to the formation of the gas giant planet around \cygb. However, the effect is not related to enhanced lithium depletion in the photosphere of \cygb\ but to the fact that the planet around \cygb\ retained a certain amount of lithium that would have otherwise been accreted by the star.

As shown in Fig.~\ref{f:trends}, the \cyga--B abundance differences do not correlate with dust condensation temperature as in M09. If we force the volatiles to match the mean abundance pattern of solar twins, the refractories are too low by 0.06\,dex and vice-versa (see solid lines in Fig.~\ref{f:trends}). If the metallicity difference between \cyga\ and B is due to the formation of the planet around \cygb, we must assume that the abundance ratios of metals in this planet are very similar to those in the star \cygb. In particular, we have to assume that the volatile to refractory abundance ratios are equal to those in the photosphere of \cygb\ (excluding H and He, of course). Thus, the planet must be a gas giant, as indeed observed. 

We explore our hypothesis quantitatively with a toy model. If the planet around \cygb\ were to be added to the convective zone of the star, the stellar metallicity would change by:
\begin{equation}
\Delta\mathrm{[M/H]}=\log\left[\frac{(Z/X)_\mathrm{CZ}M_\mathrm{CZ}+(Z/X)_\mathrm{p}M_\mathrm{p}}{(Z/X)_\mathrm{CZ}M_\mathrm{CZ}}\right],
\label{eq:dmh}
\end{equation}
where $(Z/X)_\mathrm{CZ}$ is the ratio of the fractional abundance of metals ($Z$) relative to hydrogen ($X$) in the unperturbed convective zone (loosely called ``metallicity'' in this context), $M_\mathrm{CZ}$ the mass of the convective zone, $(Z/X)_\mathrm{p}$ the metallicity of the planet, and $M_\mathrm{p}$ the mass of the planet. According to \cite{asplund09:review}, $(Z/X)_\mathrm{CZ}^\odot=0.018$. Thus, for \cygb\ ([M/H]=0.061) we adopt $(Z/X)_\mathrm{CZ}=10^{0.061}\times0.018=0.021$.

\begin{figure}
\includegraphics[bb=75 370 560 705,width=8.7cm]{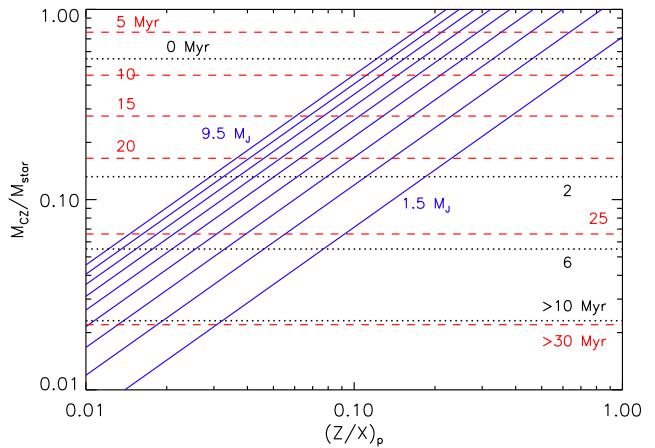}
\caption{Solid lines correspond to the mass of the convective envelope of \cygb\ necessary to explain the metallicity difference between \cyga\ and B as a function of the metallicity of the planet around \cygb\ and for planet masses between 1.5 and 9.5\,$M_\mathrm{J}$ in increments of 1.0\,$M_\mathrm{J}$. Dashed lines correspond to the mass of the convective envelope of \cygb\ for stellar ages between 5 and $>30$\,Myr according to the standard model by \cite{serenelli11}. Dotted lines represent a similar evolution for $M_\mathrm{CZ}$, from 0 to 10\,Myr, but according to one of the non-standard models with episodic accretion by \cite{baraffe10}.}
\label{f:pollution_model}
\end{figure}

We can use Eq.~\ref{eq:dmh} to calculate the mass of the convective envelope necessary to explain the observed metallicity difference of $\Delta\mathrm{[M/H]}=+0.04$ given the planetary parameters $(Z/X)_\mathrm{p}$ and $M_\mathrm{p}$. Since a minimum mass estimate of $1.5M_\mathrm{J}$ is available from the work by \cite{cochran97}, we explore Eq.~\ref{eq:dmh} for a range of planet masses from 1.5 to 9.5\,$M_\mathrm{J}$. The results from our toy model are illustrated in Fig.~\ref{f:pollution_model}.\footnote{Our simple model assumes instantaneous accretion of metal-deficient material, therefore implying that the planet itself formed instantaneously, which is a very crude approximation. More detailed models should consider the finite formation time scale of the planet in addition to the temporal evolution of the mass of the convective envelope but that is beyond the scope of our work. Our results are therefore only useful as first order approximations.}

This toy model is intended to counteract the immediate aftermath of the planet formation process. Thus, the mass of the convective envelope that appears in Eq.~\ref{eq:dmh} corresponds to the convective zone at the time of planet formation. Standard models of early stellar evolution suggest that solar analog stars begin fully convective and only after about 30\,Myr their convective envelopes reach their present-day masses \cite[e.g.,][]{dantona94,serenelli11}. On the other hand, observations of disks around young stars suggest a shorter timescale for the process of planet formation, of order 10\,Myr \cite[e.g.,][]{mamajek09,meyer09,fedele10}. Thus, we investigate our simple model keeping $M_\mathrm{CZ}$ as a variable quantity but constraining its temporal evolution as computed by \citet[][their Fig.~1]{serenelli11}. The latter corresponds to a 1\,$M_\odot$ star of solar-metallicity. To take into account the small differences in mass and $\feh$ for \cygb, we scaled the \cite{serenelli11} result using the present-day convective envelope masses as a function of $\teff,\feh$ by \cite{pinsonneault01}. The dashed lines plotted in Fig.~\ref{f:pollution_model} correspond to the mass of the convective zone at different stellar ages according to standard models. In this figure we also show, with dotted lines, the time evolution of $M_\mathrm{CZ}$ for one of the non-standard models with episodic accretion by \cite{baraffe10}. We discuss this case later in this section.

According to our pollution model with standard $M_\mathrm{CZ}$ evolution, if the planet has $(Z/X)_\mathrm{p}=0.1$ and $M_\mathrm{p}=1.5\,M_\mathrm{J}$, it should have formed when the star was about 25\,Myr old but if the planet has $M_\mathrm{p}=9.5\,M_\mathrm{J}$, it should have formed 10\,Myr after the star was born, when the convective envelope was more massive, therefore diluting more the larger amount of metal-depleted gas accreted. The planet around \cygb\ could not have formed 25\,Myr after the stellar birth since the convective envelope at that point was too small. Only if the metallicity of the planet is significantly lower ($<0.03$) it could have formed at these late times. On the other hand, if the planet is more metal rich, for example $(Z/X)_\mathrm{p}=0.7$, it must have formed in the first 10\,Myr since the birth of the star, and earlier for larger planet masses. Such large planet metallicity appears, however, unphysical for gas giants. The amount of metals in Jupiter (core and envelope) is about 12--37\,$M_\oplus$ \citep{guillot05,fortney10} and it seems that all giant planets have at least 10--15\,$M_\oplus$ of metals \citep{miller11}. Since Jupiter's mass is 318\,$M_\oplus$, a total amount of metals of 12-37\,$M_\oplus$ would imply $(Z/X)_\mathrm{p}=0.04-0.12$.

Theoretical models of planet formation predict that gas giant planets form relatively quickly, on time scales shorter than 10\,Myr \cite[e.g.,][]{pollack96,guilera11}. Yet, for reasonable values of the planet's metallicity and mass, planet formation time scales from our pollution model with standard $M_\mathrm{CZ}$ evolution are in the 10-30 Myr range, i.e., we require the process of gas giant planet formation to be significantly longer. The exact same problem is faced by M09 scenario for the formation of terrestrial planets. They have to assume that while the terrestrial planets were being formed, the mass of the solar convective envelope had already reached its present size. A possible solution to this problem may lie in the new developments in the field of early stellar evolution. In particular, the impact of episodic accretion on the internal structure of stars may help solving this discrepancy.

Models for the early evolution of stars with non-steady accretion have been computed by \cite{baraffe10}. They demonstrate that the internal evolution of stars is highly dependent on the accretion history. Of particular interest for our work is the fact that their accretion models predict higher central temperatures and hotter overall structures leading to a more rapid formation of the radiative core in solar-type stars. In one of their models, the mass of the convective envelope of a 1\,$M_\odot$ star reaches its present-day size after only about 10\,Myr from the stellar birthline. The evolution of $M_\mathrm{CZ}$ for this model in particular is represented with the dotted lines in Fig.~\ref{f:pollution_model}.

One possibility to explain the metallicity difference between \cyga\ and B in the context of gas giant planet formation, and at the same time remain compatible with the terrestrial planet signature hypothesis by M09, consists in assuming that the gas giants formed during the first $\simeq5$\,Myr after the birth of the star whereas the terrestrial planets formed later ($\gtrsim10$\,Myr). More specifically, the timescales of accretion of metal-deficient and fractionated gas should be consistent with this assumption. In addition, the evolution of $M_\mathrm{CZ}$ should be quicker than predicted by standard models; i.e., $M_\mathrm{CZ}$ should reach its present day mass in about 10\,Myr, as it is the case of some of the \cite{baraffe10} models.

\subsection{Elemental abundances in the solar nebula}

An interesting consequence of the proposed scenario is that the solar nebula would have been more metal-rich than the present-day solar photosphere. In fact, we could re-construct the original solar abundances if we had a good knowledge of the composition of the gas giants. The combined mass of the gas giants in the solar system is about $1.4\,M_\mathrm{J}$, slightly lower than the minimum mass of the planet around \cygb. On the other hand, the convective envelope of the Sun is somewhat less massive than that of \cygb. Given the uncertainties in the many parameters involved to calculate the metallicity shift due to the formation of the gas giants in the solar system, we will assume for simplicity that their impact was similar to the case of \cygb. Thus, the formation of the gas giants in the solar system could have shifted the solar abundances of all elements down by about 0.04\,dex. Later, the formation of rocky planets lowered even more the abundances of refractory elements by up to 0.08\,dex while keeping the volatile abundances nearly constant (cf.\ M09, R09). Thus, for example, the abundances of some important elements in the solar nebula would have been $\mathrm{[C,N,O/H]}\simeq+0.05$, $\feh\simeq+0.09$, and $\mathrm{[Al,Sc,Ti/H]}\simeq0.12$. Note that in this calculation we have ignored the impact of diffusion, which would further decrease the present-day observed solar abundances by about 0.05\,dex \citep{turcotte02}. Naturally, if the mass of the planet around \cygb\ is significantly larger than 1.5\,$M_\mathrm{J}$, the impact of the formation of Jupiter on the solar abundances would be lower than 0.04\,dex.

The elemental abundances measured in CM chondrite meteorites, which are often assumed to have retained the solar nebula composition, agree very well with those measured in the present-day solar photosphere \citep[e.g.,][]{lodders03,asplund09:review}. The impact of planet formation and diffusion on the solar abundances described above is not necessarily in contradiction with this observation because when these comparisons are made the
meteoritic abundances have to be normalized using the solar abundance
of Si (or a mixture of several metals with low abundance uncertainties), therefore removing any overall metallicity offsets. The trend with condensation temperature, however, would not disappear in this way but it has not yet been detected, probably owing to the still relatively large errors in the determination of absolute solar elemental abundances.

According to our proposed scenario, the differences between the initial and present-day solar abundances are not negligible and they could be relevant for example in studies of Galactic chemical evolution, chemical tagging of dynamical groups, and stellar interior modeling. Further work on the impact of planet formation on stellar abundances, both from the observational and theoretical sides, is therefore urgently needed.

\section{Summary}

We have determined the stellar parameters and abundances of 25 elements in \cyga\ and \cygb\ with very high precision. Their lithium abundances appear perfectly normal considering the observed lithium depletion patterns in field solar-type stars as a function of mass and metallicity and the small (but non-negligible) differences in these two parameters for the 16\,Cyg stars. Accretion of planetary material onto \cyga\ and enhanced photospheric lithium depletion due to the presence of the planet around \cygb\ are not necessary to explain the observed lithium abundances.

Excluding lithium, we show that \cyga\ is more metal-rich than \cygb\ by $\Delta\mathrm{[M/H]}=+0.041\pm0.007$\,dex. The abundance differences detected do not correlate with dust condensation temperature. Thus, if the difference is due to the accretion of metal-depleted gas onto \cygb, the abundance ratios of different metals in this gas, in particular the volatile to refractory abundance ratio, should be nearly the same as that found today in the photosphere of \cygb. We propose a scenario in which both \cyga\ and B started with the exact same composition but the formation of a planet around \cygb\ made this secondary star slightly metal-poor. The composition of the planet must be the same as that of the stars (with the exception of H and He) in order for this hypothesis to work on a quantitative basis. Moreover, the planet must have formed while the convective envelope of the star was still relatively massive (about 10\,\% of the stellar mass).

The idea proposed to explain the abundance differences in the 16\,Cyg binary system is similar to that suggested by \cite{melendez09:twins} and \cite{ramirez09} to explain the anomalous solar abundances as a signature of the formation of terrestrial planets. The main difference is that the planet around \cygb\ is a gas giant. We can reconcile these two hypothesis by assuming that gas giants form while the convective envelope is massive whereas terrestrial planets form later, when the convective zone reaches its present-day size. In any case, the formation of the stellar radiative core should be quicker than that predicted by standard stellar evolution models. The impact of episodic accretion on early stellar evolution could provide a solution to this problem.

If this scenario is correct, we must conclude that the Sun's original metallicity was higher, an effect possibly enhanced by atomic diffusion. Perhaps the regions immediately below the base of the solar convective envelope are significantly more metal-rich than the photosphere, a condition that has been proposed as a possible solution to the so-called solar abundance crisis \citep[e.g.,][]{nordlund09}, i.e., the severe discrepancy between the observed photospheric solar abundances and those inferred from helioseismological data and stellar interior models \citep[e.g.,][]{basu04,basu08,bahcall05,delahaye06}. Note, however, that this idea seems not to work very well on a detailed, quantitative basis \cite[e.g.,][]{serenelli11}.

Our detail study of the 16\,Cyg system shows that high precision elemental abundance work in wide binary stars with planets can provide us with important clues to understand the impact of planet formation on the chemical composition of stars. We encourage future spectroscopic studies of binary stars with similar components, both to measure small differences in their chemical composition and to look for planets, especially if the stars are solar analogs.

\acknowledgments

I.R.'s work was performed under contract with the California Institute of Technology (Caltech) funded by NASA through the Sagan Fellowship Program. J.M. would like to acknowledge support from FAPESP (2010/17510-3). I.U.R.\ is supported by the Carnegie Institution of Washington through the Carnegie Observatories Fellowship. We thank P.~Barklem for computing the fine grid of theoretical H$\alpha$ profiles used to measure $\teff$. I.R.\ would also like to thank D.\ L.\ Lambert, D.~R.~Doss, and the McDonald Observatory staff for observing support.

\end{document}